\documentclass{aastex}

\usepackage{epsfig}
\usepackage{amsmath}
\usepackage{color}
\usepackage{subfigure}
\usepackage{textcomp}
\def\ed{\end{document}}

\def \kms {km s$^{-1}$}

\begin{document}
\title{The Nature of the Density Clump in the Fornax Dwarf Spheroidal Galaxy}
\author{Edward W. Olszewski \affil{Steward Observatory, University of Arizona, Tucson, AZ 85721}\email{eolszewski@as.arizona.edu}}
\author{Mario Mateo \affil {Department of Astronomy, University of Michigan, 830 Dennison Building, Ann Arbor, MI 48109-1090}\email{mmateo@umich.edu}}
\author{Jason Harris \affil {Steward Observatory, University of Arizona, Tucson, AZ 85721}\email{jharris@as.arizona.edu}}
\author{Matthew G. Walker \affil{Department of Astronomy, University of Michigan, 830 Dennison Building, Ann Arbor, MI 48109-1090}\email{mgwalker@umich.edu}}
\author{Matthew G.~Coleman and G.~S.~Da~Costa\affil{Research School of Astronomy and Astrophysics, Institute of Advanced Studies, Australian National University, Cotter Road, Weston Creek, ACT 2611, Australia}\email{coleman@mso.anu.edu.au,gdc@mso.anu.edu.au}}
\begin{abstract}
We have imaged the recently discovered stellar overdensity located approximately one
core radius from the center of the Fornax dwarf spheroidal
galaxy using the Magellan Clay 6.5m  telescope with the Magellan Instant Camera (MagIC). Superb
seeing conditions allowed us to probe the stellar populations of this overdensity
and of a control field within Fornax to a limiting magnitude of R=26.
The color-magnitude
diagram of the overdensity field is virtually identical to that of the control
field with the exception of the presence of a population arising from a very short (less
than 300 Myr in duration) burst of star formation
1.4 Gyr ago. 
Coleman et al.~have argued that this overdensity might be related to
a shell structure in Fornax that was created when Fornax captured a smaller
galaxy. 
Our results are consistent with this model, but we argue that the
metallicity of this young component favors a scenario in which
the gas was part of Fornax itself.

\end{abstract}

\keywords {galaxies: dwarf, galaxies: individual (Fornax dSph), Local Group,
galaxies: stellar content, Magellan: Clay}

\section{Introduction}

Our understanding of the complexity of the star formation histories of
dwarf spheroidal (dSph) galaxies has advanced with each new telescope
and detector improvement.  Dwarf spheroidal galaxies were first
thought to have the same stellar populations, and total masses, as
globular clusters, with the only difference being the larger physical
size of the spheroidals.  The discovery of carbon stars in these
systems (Westerlund 1979; Demers \& Kunkel 1979; Aaronson \& Mould     
1980; Cannon et al.~1981; Aaronson et al.~1982, 1983), and
the realization that the peak luminosity of populations of C stars was an
age indicator (Mould \& Aaronson 1979; Aaronson \& Mould 1980; Aaronson \& Mould 1985 and references therein)
showed that dwarf spheroidals had, in fact,  much more complicated star formation
histories. 
The first color-magnitude diagram to reach the main sequence of a
dSph (Mould \& Aaronson 1983) spectacularly confirmed the youth of a major
fraction of the stars in Carina. 
Subsequent observations have shown that other dSph's have
diverse populations:
some have large or dominant populations significantly younger
than those of Galactic globulars, for instance,
Carina (Mould \& Aaronson 1983,
Smecker-Hane et al.~1996; Hurley-Keller et al.~1998 ), Leo~I (Gallart
et al.~1999a, 1999b), and Fornax (Stetson et al.~1998, Saviane et
al.~2000) (see the schematic star formation histories in Mateo 1998).
Others have less pronounced intermediate-age populations, while
Ursa Minor (Olszewski \& Aaronson 1985), with almost 100\% ancient
stars, is the closest to a globular-like population.
Given the physical properties of dSph galaxies,
even star formation of the first population of stars in dwarf
spheroidals is difficult to understand, but the problem of multiple
generations of stars has led to several potential solutions for
quenching or restarting star formation (Lin \& Faber 1983; Silk et al.~1987; Burkert
\& Ruiz-Lapuente~1997; Mashchenko et al.~2004).

Dwarf spheroidals not only have complex star formation histories, but
are more complex than globulars structurally, with radial profiles
(from star counts) that are not well fit by simple one-component King
models.  Besides this global structural difference, some dSph's have
demonstrable asymmetries (Stetson et al.~1998) or substructure.
Olszewski \& Aaronson (1985) showed that the distribution of stars at
the center of Ursa Minor was not smooth. This Ursa Minor ``cluster''
has been rediscovered several times (e.g., Demers et al.~1995), along
with other star-count bumps in UMi (Demers et al.~1995, Kleyna et
al.~2003) and similar structures in other dwarf spheroidals (Eskridge
1988a,b; Kleyna et al.~2004). Recently, it has been argued that some
of these density enhancements have kinematic properties that differ from
those in typical fields in these dSph's (Kleyna et al.~2003, 2004).
Certainly, some special property of these higher stellar density
regions seems necessary, for one or both of the dynamical mixing times
and the crossing times,$\sim 10^8$ yr,  are short compared to the age of the stars in
the clump, $\sim 10^9$ yr.

  Coleman et al.~(2004, 2005) have recently
probed substructure in Fornax, showing that a density enhancement (hereafter
known as the ``clump'') at
about the core radius of Fornax contained a population of
younger-than-average stars.  They have also shown that Fornax may have
shells rather like those in giant ellipticals, speculating that Fornax
itself suffered a collision with a smaller halo system. This collision
created the outer 
shells discovered in Coleman et al.~(2005) and led to star
formation that generated the clump in Fornax. In this way Fornax
may have served as a ``detector'' to reveal the existence
of a low-mass halo system, possibly gas-rich and star-free,
that otherwise may have remained invisible. The Local Group is
apparently deficient in small halos when compared to simulations of
the formation of structure (e.g., Moore et al., 1999, Klypin et al., 1999).

Since Fornax has no neutral gas today (Knapp et al.~1978, Young 1999), one
class of solutions to a fairly recent burst of star formation is to speculate
a gas-rich collider. This collider could equally well be a hydrogen
cloud, a smaller version of the sort recently discovered by Minchin et
al.~(2005), or a high-velocity cloud.  We of course would want to know
how much gas, and the etymology of that gas, there was in Fornax
during the actual star formation episodes, but aside from counting
stars of various ages and assuming a star formation efficiency, there
seems to be no way to know the gas mass at various ages.  Knapp et
al.~(1978) quoted a very low current H I mass from observations
centered at a heliocentric velocity (badly known at the time) of 35
\kms\ $\pm$6.7 \kms, which does not overlap the mean stellar velocity
of 53 $\pm$ 3 \kms\ (compiled in Mateo 1998, Walker et al.~2005).  A
more recent measurement of H I gas in Fornax concludes that no
emission or absorption is detected to limits of about
10$^{18}$--10$^{19}$ cm$^{-2}$ (Young 1999), which corresponds to a
mass of 10$^4$--10$^5$ M$_\odot$ within the 26 arcmin  ($\sim 1$ kpc)
VLA
observation. Using a density of 10$^{19}$ cm$^{-2}$, the roughly
100$\times$100 pc stellar overdensity could be hiding roughly 2000
M$_\odot$ of potentially star-forming gas even today. This is a very
frustrating limit because it is similar in magnitude to the mass of
stars in the Fornax clump. This limit invites the opposite of the
Coleman et al.~speculation: perhaps gas did not need to be delivered
to Fornax, but was compressed and shocked into forming stars. 
Indeed it is possible
that Fornax could contain significant amounts of warm (T $\approx 10^4$ K)
ionized gas (cf.\ Gallagher et al. 2003), though Gizis et al. (1993) give
an upper limit of $\sim10^5$
M$_\odot$
 for any hot (T $\approx 10^6 - 10^7$
K) diffuse gas in Fornax.  We await more sensitive surveys of the diffuse
interstellar medium in Fornax as they may hold the key to understanding
how multiple generations of star formation occur in dSphs.
\section{Observations and Data Reduction}

The photometric observations were carried out on Dec 22 and 23 (UT),
2003 using the Magellan Instant Camera (MagIC, see Osip et al.~2004 for
a brief description) on the Magellan/Clay
Telescope at Las Campanas Observatory. While we observed at Las
Campanas with another instrument, problems with that instrument caused
us to switch to the MagIC Camera, which is always mounted and running,
and available with a shift of the tertiary mirror. This ability saved
two excellent nights, and allowed us to probe the Coleman et al.~(2004)
clump to unprecedented depth.  MagIC contains a single $2048 \times
2048$ SITe CCD with 24 micron pixels.  This corresponds to a scale of
0.069 arcsec per unbinned pixel.  Unless noted, all of the exposures
used in this study were read from the CCD unbinned.  For this project
we used a Johnson B filter and a Kron-Cousins R filter
(see Sung \& Bessell (2000), and references therein to older Bessell
publications, for a definition of the characteristics of these
passbands). Each field was observed on both nights.

Seeing was superb on these nights, averaging 0.3-0.4 arcsec.  As both
nights were photometric, we obtained data for Stetson (2000)\footnote{
see the URL http://cadcwww.hia.nrc.ca/standards for updated lists of
standard fields and stars.} standards throughout the run.  The Stetson
fields extend Landolt (1983, 1992) fields and also provide new fields,
for instance, stars in Fornax itself. These standard-field raw data
were binned $2 \times 2$ at the telescope during readout to speed up
the calibration observations.  We measured all standard-star images
using IRAF/qphot\footnote{IRAF is the Image Reduction and Analysis
Facility, a general purpose software system for the reduction and
analysis of astronomical data.  IRAF is written and supported by the
IRAF programming group at the National Optical Astronomy Observatories
(NOAO) in Tucson, Arizona.  NOAO is operated by the Association of
Universities for Research in Astronomy (AURA), Inc. under cooperative
agreement with the National Science Foundation.} in apertures of 10,
12, and 15 pixels radius, correcting all 10 pixel photometry to 15
pixel photometry (4 arcsec diameter, a factor of 10 larger than the
seeing in the program fields) using a single aperture correction per
field.  The initial transformations of the measured instrumental
magnitudes and colors to the calibrated magnitudes and colors were
essentially identical for both nights.  We therefore decided to
combine all of the standards from both nights into one dataset from
which we determined one transformation.  The resulting transformation
equations are $$R = 1.946 + r - 0.180X + 0.022(b-r),$$ and $$(B-R) =
-0.221 - 0.198X + 1.036(b-r),$$ where X is the airmass of
observation. The lower-case terms represent the instrumental values
(corrected to one-second exposure time) and the upper case are the
values as taken from Stetson's online catalog. Standard deviations of
individual standard-star observations about these equations are 0.017
mag for magnitude and 0.016 mag for color.

Two fields were observed in the Fornax dSph galaxy.  One was centered
on the local overdensity identified by Coleman et al.~(2004) and the
other was positioned at an approximately diametrically opposite
position relative to the center of Fornax.  The latter is our
`control' field.  For the principal Fornax clump field we obtained nine
900-sec B images and nine 600-sec R images. For the control field we
obtained nine 900-sec B images and eight 600-sec R images.  Shorter
exposures, of length 30--90 sec, were also obtained in both
fields. The individual long-exposure images were shifted to a common
coordinate system (using only integer pixel shifts), and combined
using a sigma-clip to remove the effects of cosmic rays.  These
combined images, with a total exposure of 2.25 hours in B and 1.5
hours in R for the clump field and 2.25 hours in B and 1.33 hours in R
for the control field, have full-widths-at-half-maximum between 0.4
and 0.5 arcsec in both filters and fields.  A summary of all exposures
is given in Table 1.  Figures 1 and 2 are images of the clump and
control fields. Tables 2 and 3, which contain coordinates and
photometry, are truncated in the print edition of this paper; the full
Tables are available in the electronic version.  RA and Dec are
reported from coordinates derived using IRAF tasks TFINDER and CCTRAN
using approximately 50 USNO-B1 stars (Monet et al.~2003\footnote{We
used the online data available at http://www.nofs.navy.mil/ }) in
each 2.3$\times$2.3 arcmin field; the solutions to the standard
stars had a scatter of about 0.2 arcsec in each coordinate.
Photometry is excellent to R=26 mag; typical internal psf-fitting
errors of the artificial stars (section 2.1) are 0.01-0.015 mag to
R=21.5, 0.03-0.04 mag at R=23-25, and 0.05-0.06 mag at
R=25.5-26.5. The resultant color-magnitude diagrams are reminiscent of
WFPC2 data of similar exposure time (for example Gallart et
al. 1999a), and show the strengths of ground-based imaging on modern,
large telescopes designed to deliver excellent image quality.

\begin{figure}
 \begin{center}
  \plotone{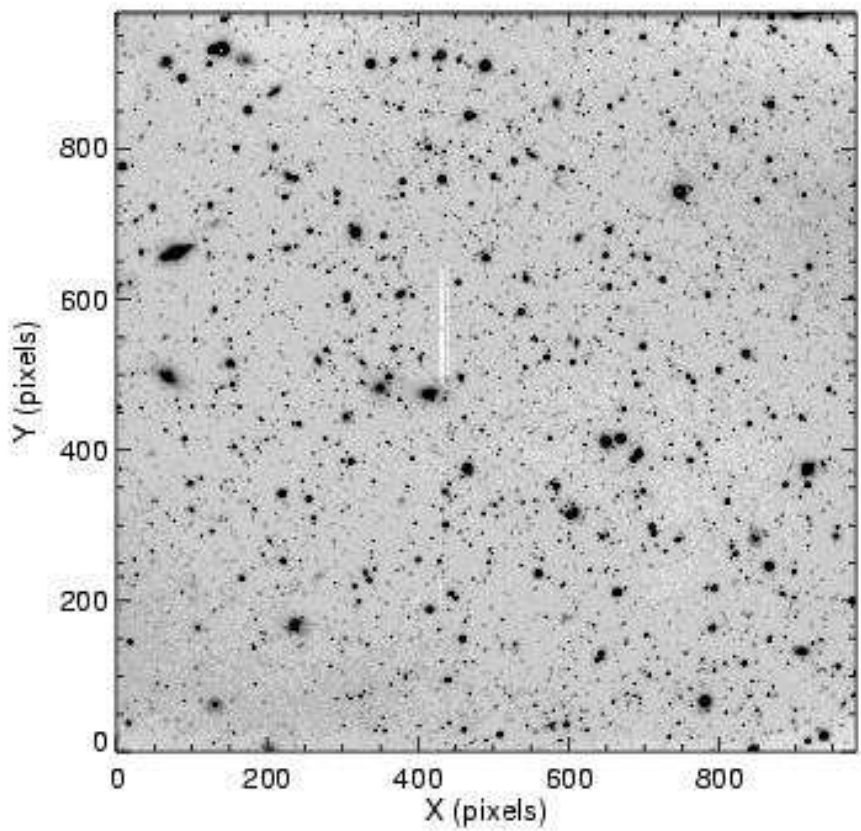}
   \caption{\label{fig:cmds1} \scriptsize The R-band image
of the Fornax clump field. It is the average (with cosmic ray
rejection) of nine 600-sec individual images. The FWHM of
this combined image, which was binned 2$\times$2 for the photometry, is between
0.4 and 0.5 arcsec. The xy scale refers to binned coordinates in Table 2;
stars can be most easily identified for future work by their (RA,Dec).
The central coordinates of this field are (RA) 02:40:28.0 (2000); (Dec) -34:42:33 (2000).}
 \end{center}
\end{figure}

\begin{figure}
 \begin{center}
  \plotone{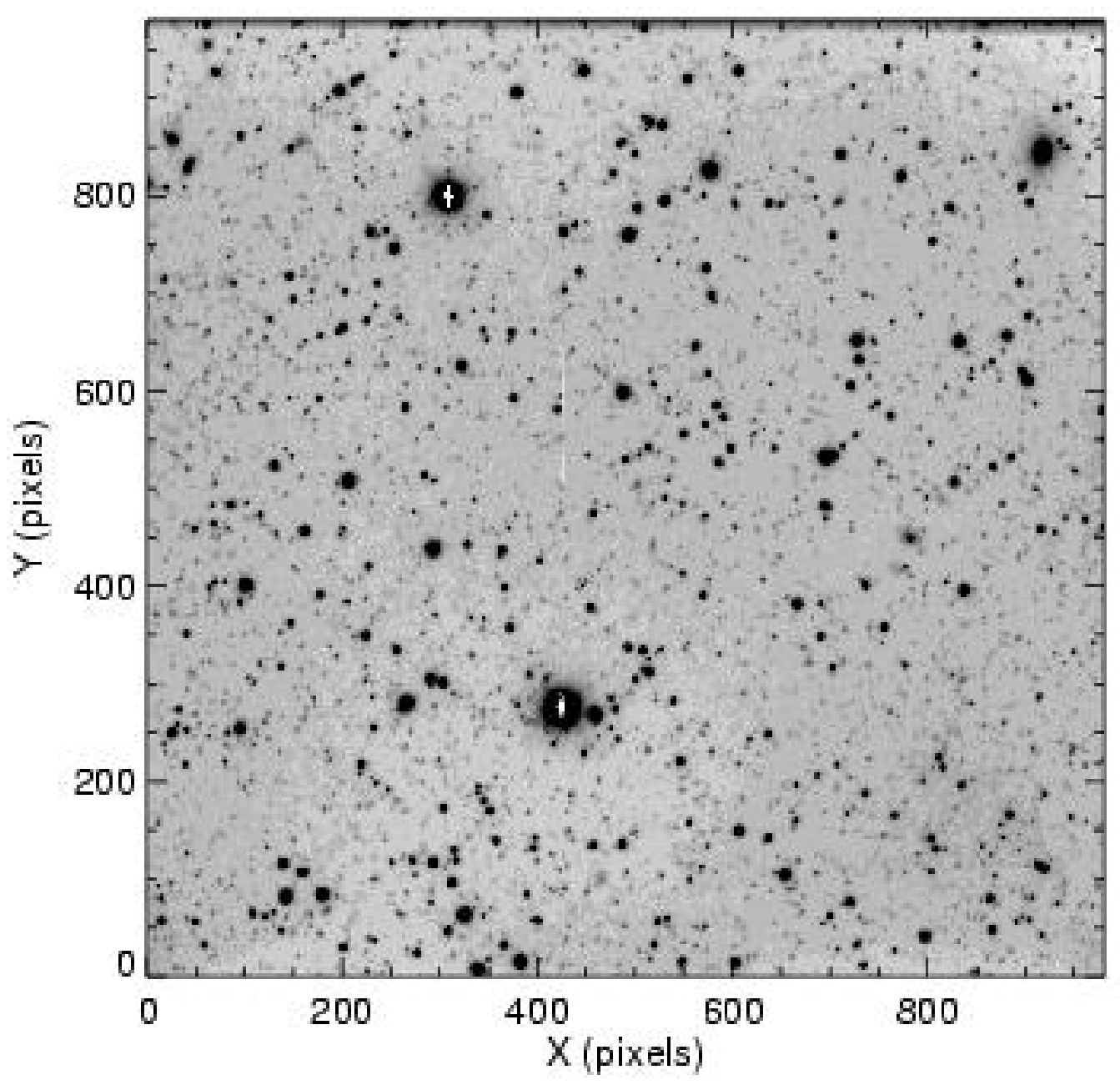}
   \caption{\label{fig:cmds2} \scriptsize The R-band image
of the Fornax control field. It is the average (with cosmic ray
rejection) of eight 600-sec individual images. The FWHM of
this combined image, which was binned 2$\times$2 for the photometry, is between
0.4 and 0.5 arcsec. The xy scale refers to binned coordinates in Table 3;
stars can be most easily identified for future work by their (RA,Dec).
The central coordinates of this field are  (RA) 02:39:18.4 (2000); (Dec) -34:21:37.5 (2000).}
 \end{center}
\end{figure}

\begin{deluxetable}{lcrccl}
  \tablenum{1}
  \tabletypesize{\scriptsize}
  \tablewidth{0pc}
  \tablecaption{\scriptsize Table of Observations
  \label{tab:log}}
\tablehead{\colhead{Field\tablenotemark{a}} & \colhead{Filter} &  
\colhead{Date (UT)} & \colhead{Exposure} & \colhead{$N$\tablenotemark{b}}&
\colhead{Airmass}\\
\colhead{}&\colhead{}&\colhead{}&\colhead{Time (sec)}&\colhead{}&\colhead{Range} }
  \startdata
Fornax Clump &B&2003/12/22-23&900&9&1.01-1.13\\
Fornax Clump &R&2003/12/22-23&600&9&1.01-1.10\\
Fornax Clump &B&2003/12/22&30&1&1.02\\
Fornax Clump &B&2003/12/23&90&3&1.01\\
Fornax Clump &R&2003/12/22&30&1&1.01\\
Fornax Clump &R&2003/12/23&60&3&1.01\\
Fornax Control &B&2003/12/22-23&900&9&1.19-1.77\\
Fornax Control &R&2003/12/22&600&8&1.15-2.01\\
Fornax Control &B&2003/12/23&90&3&1.12\\
Fornax Control &R&2003/12/23&60&3&1.12-1.13\\
Stetson Fornax 983 field &B&2003/12/22-23&30&12&1.10-2.09\\
Stetson Fornax 983 field &R&2003/12/22-23&12&12&1.10-2.13\\
Stetson E5 S72 field &B&2003/12/23&30&3&1.14\\
Stetson E5 S72 field &R&2003/12/23&12&2&1.13\\
Stetson NGC2818-S92 field &B&2003/12/22-23&30&6&1.01\\
Stetson NGC2818-S92 field &B&2003/12/23&20&3&1.01\\
Stetson NGC2818-S92 field &R&2003/12/22&12&6&1.01\\
Stetson NGC2818-S92 field &R&2003/12/23&10&4&1.01\\
  \enddata
\tablenotetext{a}{``Stetson'' refers to standard fields available at CADC.}
\tablenotetext{b}{total number of images}
\end{deluxetable}

\begin{deluxetable}{lcccccccc}
  \tablenum{2}
  \tabletypesize{\scriptsize}
  \tablewidth{0pc}
  \tablecaption{\scriptsize Photometry of the Fornax Clump Field
  \label{tab:clump_phot}}
\tablehead{\colhead{Star} & \colhead{$\alpha_{2000}$}  & \colhead{$\delta_{2000}$} & \colhead{X} & \colhead{Y} & \colhead{$R$} & \colhead{$err_R$} & \colhead {$B-R$} & \colhead{$err_{B-R}$}}
  \startdata
     1 &  2:40:33.91 & -34:42:52.3&  369.58 &    1.35 &  24.513 &   0.028 & 0.603&    0.050 \\
     3 &  2:40:33.93 & -34:41:51.8&  804.74 &    1.95 &  25.176 &   0.062 & 0.643&    0.082 \\
     2 &  2:40:33.93 & -34:41:46.4&  843.74 &    1.95 &  25.190 &   0.062 & 1.325&    0.118 \\
     5 &  2:40:33.93 & -34:41:41.2&  880.87 &    2.35 &  25.750 &   0.105 & 0.372&    0.143 \\
     4 &  2:40:33.88 & -34:43:28.6&  109.41 &    2.35 &  25.528 &   0.189 & 0.659&    0.220 \\
     6 &  2:40:33.90 & -34:42:56.4&  340.41 &    2.39 &  23.945 &   0.030 & 0.513&    0.036 \\
     7 &  2:40:33.88 & -34:43:23.2&  148.11 &    2.80 &  22.976 &   0.017 & 1.050&    0.028 \\
     8 &  2:40:33.91 & -34:42:03.5&  720.28 &    3.04 &  25.661 &   0.103 & 0.950&    0.155 \\
    11 &  2:40:33.91 & -34:42:01.6&  733.79 &    3.36 &  23.091 &   0.022 & 0.240&    0.024 \\
    13 &  2:40:33.90 & -34:42:32.1&  514.59 &    3.59 &  24.247 &   0.043 & 0.721&    0.059 \\
  \enddata
\end{deluxetable}

\begin{deluxetable}{lcccccccc}
  \tablenum{3}
  \tabletypesize{\scriptsize}
  \tablewidth{0pc}
  \tablecaption{\scriptsize Photometry of the Fornax Control Field
  \label{tab:control_phot}}
\tablehead{\colhead{Star} & \colhead{$\alpha_{2000}$}  & \colhead{$\delta_{2000}$} & \colhead{X} & \colhead{Y} & \colhead{$R$} & \colhead{$err_R$} &\colhead {$B-R$} & \colhead{$err_{B-R}$}}
  \startdata
     2 &   2:39:25.20 & -34:21:10.2 &640.99 &    0.21 &  23.271 &   0.012& 1.206 &   0.022\\
     3 &   2:39:25.21 & -34:20:46.7 &810.26 &    0.24 &  21.768 &   0.019& 1.431 &   0.033\\
     4 &   2:39:25.19 & -34:21:04.8 &680.36 &    0.87 &  24.463 &   0.035& 0.731 &   0.053\\
     5 &   2:39:25.19 & -34:21:04.8 &680.33 &    0.89 &  24.453 &   0.035& 0.742 &   0.053\\
     6 &   2:39:25.19 & -34:21:04.7 &680.53 &    0.90 &  24.371 &   0.050& 0.828 &   0.064\\
     7 &   2:39:25.18 & -34:22:13.2 &185.28 &    0.98 &  24.435 &   0.037& 0.675 &   0.065\\
    11 &   2:39:25.19 & -34:20:55.4 &747.45 &    1.52 &  22.566 &   0.006& 0.754 &   0.011\\
    12 &   2:39:25.19 & -34:20:54.6 &753.50 &    1.69 &  18.796 &   0.006& 1.821 &   0.411\\
    13 &   2:39:25.17 & -34:21:20.6 &566.36 &    1.84 &  25.196 &   0.069& 0.794 &   0.107\\
    14 &   2:39:25.17 & -34:21:23.1 &548.37 &    1.87 &  20.968 &   0.006& 1.201 &   0.009\\
  \enddata
\end{deluxetable}

\subsection{Photometry and False-Star Tests}

The photometry of both Fornax fields was measured using DoPHOT
(Schechter et al.~1993), a psf-fitting routine used to obtain
precise photometry with minimal user intervention and at high
computation speed.  All frames were binned 2x2 before running DoPHOT
because the full width of even 0.4 arcsec images is of order 7
pixels.
We carried out a series of false-star tests to allow us to
estimate the photometric completeness and precision empirically as a
function of position within the color-magnitude diagram (CMD).  These parameters are needed
for later analysis. Numerically-generated stars were
added to the deep B and R images at a specified position and with
specified R magnitudes and (B--R) colors.  The photometric properties
of these false stars were chosen to lie uniformly in the range $-0.5
\leq (B-R) \leq 2.5$, and $17.5 \leq R \leq 27.5$.  The locations of
the stars were chosen to lie on a regular grid in the original image
with each false star separated by 10 seeing diameters from its nearest
false neighbor.  This step maximizes the number of stars added to each
image while ensuring that false stars only affect, and are only
affected by, images of actual stars in the Fornax fields.  In adding
these stars we explicitly took into account the photometric
transformations, the position offsets between the images, the
variation of the PSF with position on the B and R images
independently, and the variations of the aperture corrections with
position and brightness on both images.  A total of 500 stars were
added to each of 50 $B,R$ pairs in each field, corresponding to 25,000
false stars for each field.  These images were then reduced with
DoPHOT in {\it precisely} the same manner as the original images.
Stars from these reductions were then matched, by position, with the
false star input positions. The injected magnitudes in B and R, and the
recovered measured magnitudes were then compared.  Stars that were not detected in {\it both}
false-star images were flagged; these cases help us define the
completeness of the photometry.

\section {Results and Conclusions}

The resulting color-magnitude diagrams for both the principal
and control fields are shown in Figure 3.  These CMDs are made
by combining the deep photometry with the shallow photometry. Approximately
6000
stars are plotted in each field.

A reddening correction of E(B$-$V)$=$0.02 (Schlegel et al.~1998) needs
to be applied to these data before extracting astrophysical
information.  In addition, to match the principal older components in
the clump and control cmds, a color offset of $+0.02$ mag was applied
to the clump field and an offset of $-0.02$ mag was applied to the
control field.  These offsets are within the errors of the photometric
solution coupled with the errors in the aperture correction from psf
magnitude to total magnitude.

Some well-known features of Fornax are visible in Fig.\ 3, including the extended
and somewhat broad (in color) giant branch, the mostly-red horizontal
branch and helium-burning clump, as well as a few additional blue horizontal branch
stars at slightly fainter mean R magnitudes.  Each diagram also shows
that most stars in these fields come from a population
with a main-sequence turnoff at $R \sim 23.5$.  The
well-defined subgiant branches emanating from that turnoff region show
significant breadth in magnitude and color, indicating that this
population likely exhibits a significant age and metallicity spread.

All of these features in the two fields match quite well with the
exception of the main sequence stars brighter than about R=23.5 mag,
the younger main sequence, extending to $R \sim 22$ from the older MS
turnoff point at $R \sim 23.5$.  Although there are such stars in
the control field -- indicative of a pervasive intermediate-age
population in that field and perhaps across Fornax (Stetson et al.~1998)-- these bluer
main-sequence stars define a clearer and sharper distribution in the
clump field.  Coleman et al.~(2004) identified these stars in their
original description of the clump, finding an age, from a rough analysis
of a Hess diagram of the clump with statistical
subtraction of the Fornax field, of approximately 2
Gyr for this younger population. Their available photometry was
insufficiently deep to allow Coleman et al.~to define this population
of stars more precisely.

The first qualitative conclusion is thus that the clump contains an
excess of younger main sequence stars, consistent with the original
analyses of Coleman et al.\ (2004, 2005). This region of
Fornax experienced significant star formation more recently than the
control field. The control field nevertheless contains young stars, indicating
that Fornax has had some star formation until recent times
(Stetson et al.~1998). 
Some of the control-field stars might be younger than any in the clump,
but we argue below that the clump contains a significant burst of
intermediate-age star formation.

\begin{figure}
 \begin{center}
  \plotone{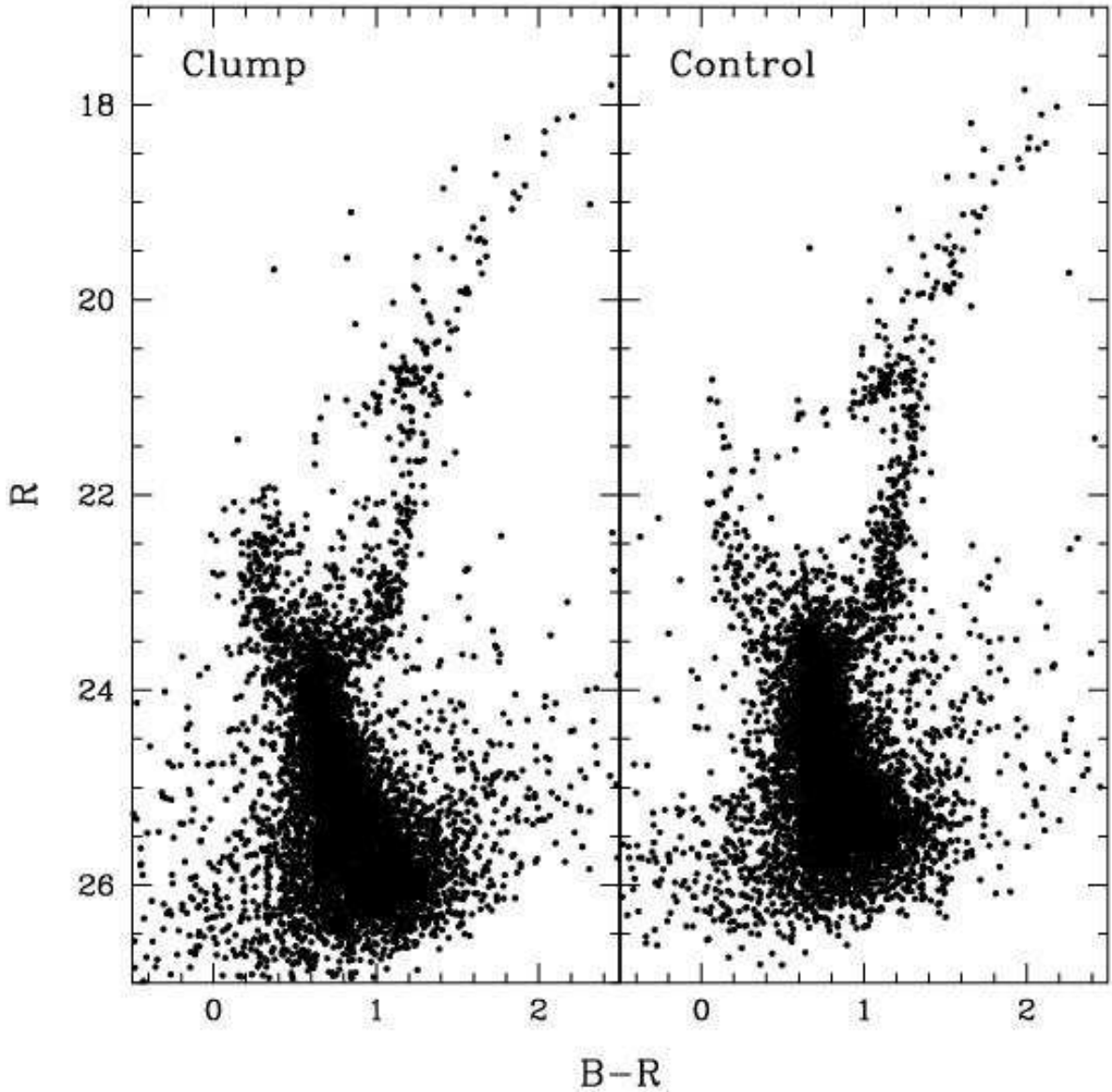}
   \caption{\label{fig:cmds} \scriptsize Color-magnitude diagrams
of the Fornax clump and control fields. Clump is to the
left, with the control field to the right. Both CMDs have
approximately 6000 stars. The control has 1.5$\times$ as many
giants, but both fields have the same number of ``young''
main-sequence stars. The clump is therefore enriched
in this younger population compared to a random Fornax field.
Were the CMDs scaled to the same number of giants and subgiants,
the additional young population in the clump would be even more apparent.}

 \end{center}
\end{figure}

The control field was chosen to lie at the diametrically opposite
point as the clump field through the center of Fornax, but  
does not lie
precisely on the same isopleth as the clump field.
Estimating from the star counts of Eskridge (1988a,b)
and of Irwin \& Hatzidimitriou (1995), we find that the control stellar
density to the limits of those two star count studies is a factor of
approximately 1.5--2.0 larger than that of the clump field.

We can test this density difference from the CMDs themselves.  A box
encompassing the entire RGB and red HB gives 255 clump stars and 378
control stars, a factor of 1.48 higher in the control. The lower SGB
gives 169 clump stars and 253 control stars, a factor of 1.49.  These
numbers are consistent with those derived from the rather coarse and
shallow isopleths. The older MSTO region gives 1597 clump stars and
2033 control stars, a factor of 1.27, while the younger main sequence
gives 412 clump stars and 398 control stars, a factor of 0.96.  These
latter two results are consistent with the fact that the clump
contains an excess of younger stars. Scaling by a density factor of 1.5,
the clump has a density of these younger stars roughly
1.5$\times$ that of the control field.

A model inspired by Coleman et al.~(2004) is that the clump is a
typical region of Fornax with the additional of a burst of star
formation at a younger than average age. This idea is qualitatively
reasonable from Figure 3 and from our own star counts.  We test this
idea more carefully by determining a quantitative star formation
history (SFH) for each field, using the StarFISH code (Harris \&
Zaritsky 2001).  StarFISH performs a statistical comparison between
the observed photometry and model photometry, derived from theoretical
isochrones and convolved with photometric errors derived from the
artificial stars tests. The real distribution of photometric errors is
used in StarFISH, not the average values given earlier in this paper.

Because the observed stellar populations contain a relatively small 
number of stars ($\sim6000$ each), we bin the CMD plane rather coarsely
($0.2\times0.2$~mag) to reduce shotnoise in the SFH solution.  We 
adopt a single, average extinction value of $A_V=0.05$~mag for both 
fields.  We use the Padua isochrones (Girardi et al.~2002), and our 
grid of synthetic CMDs covers three metallicities (Z=0.001, Z=0.002, 
and Z=0.004), and ages between 100~Myr and 15~Gyr, with logarithmic age 
bins ($\Delta log(age)=0.2$).  

The best-fit SFH solutions in each field are shown in Figure~4, in 
which we have summed over the star formation rates from our three 
metallicity bins to show the total star formation rate as a function of 
time.  In Figure~5, we compare the two SFH solutions directly, after 
scaling the clump field solution by a normalization factor of 1.5 
(see above).  While the two SFH solutions are nearly indistinguishable 
within their confidence regions, there is a small excess in the clump 
population's star formation rate at an age of 1.6 Gyr (log(age)$=$9.2).  
The confidence 
intervals are large due to the small number statistics, and 
due to the fact that they also
include correlated uncertainties between adjacent SFH amplitudes.  
We believe the actual burst of star formation in the Clump was of much 
shorter duration that the interval covered by the log(age)=9.2 bin (see 
below).  However, StarFISH is only sensitive to the average SFR over the full 
interval of each bin. Therefore the actual instantaneous SFR during the burst 
was almost certainly higher than shown in Figures 4 and 5.

\begin{figure}
 \begin{center}
  \plotone{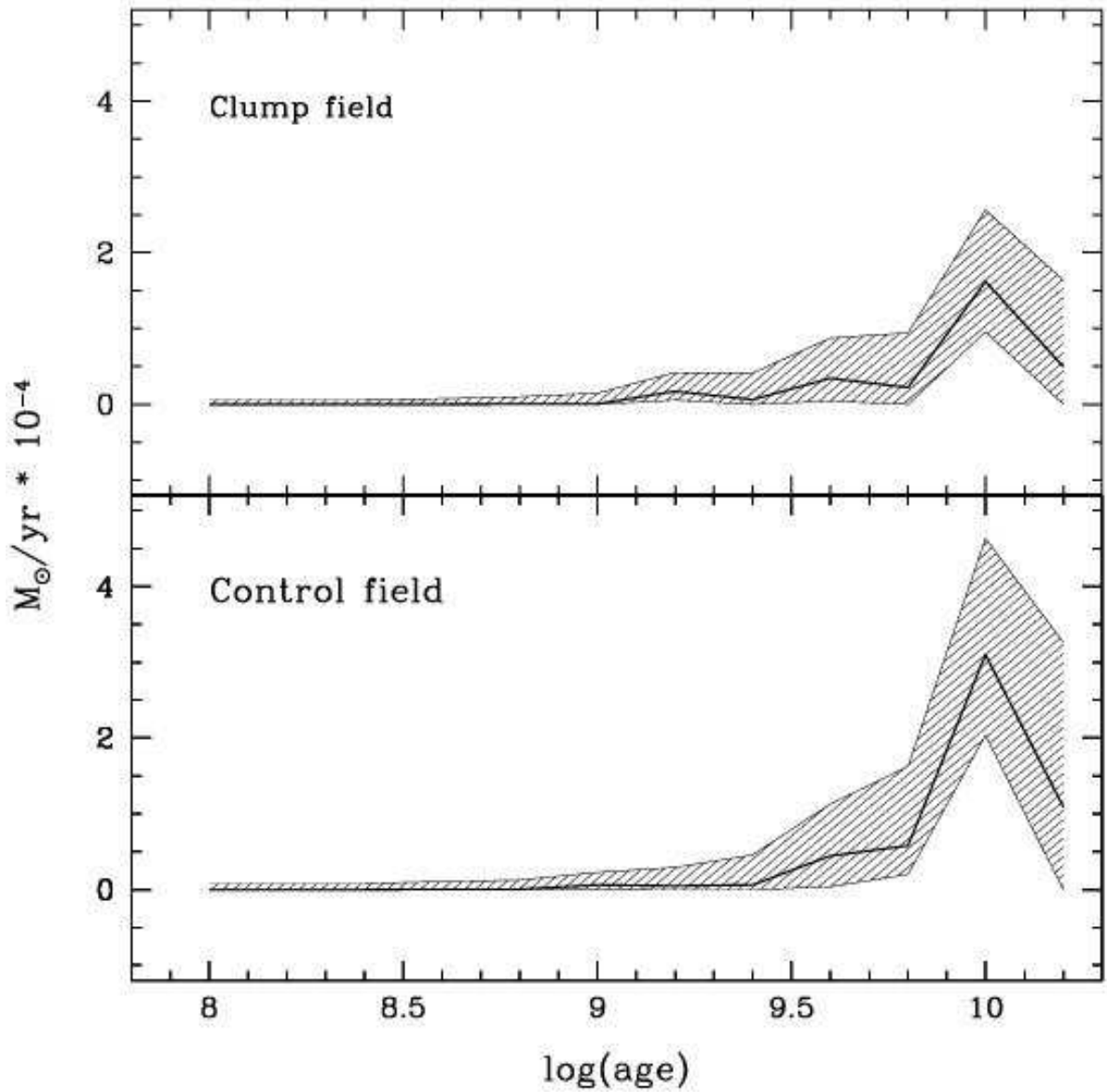}
   \caption{\label{fig:sfh} \scriptsize Total star formation history
in the clump field (top) and control field (bottom). The three
metallicity bins have been summed to give the total star-formation rate
in each age bin. The hatched area shows the 68\% confidence limits.
Reddening and attenutation corresponding to E(B-V)$=$0.02
have been applied to the isochrones before calculation.}
 \end{center}
\end{figure}

\begin{figure}
 \begin{center}
  \plotone{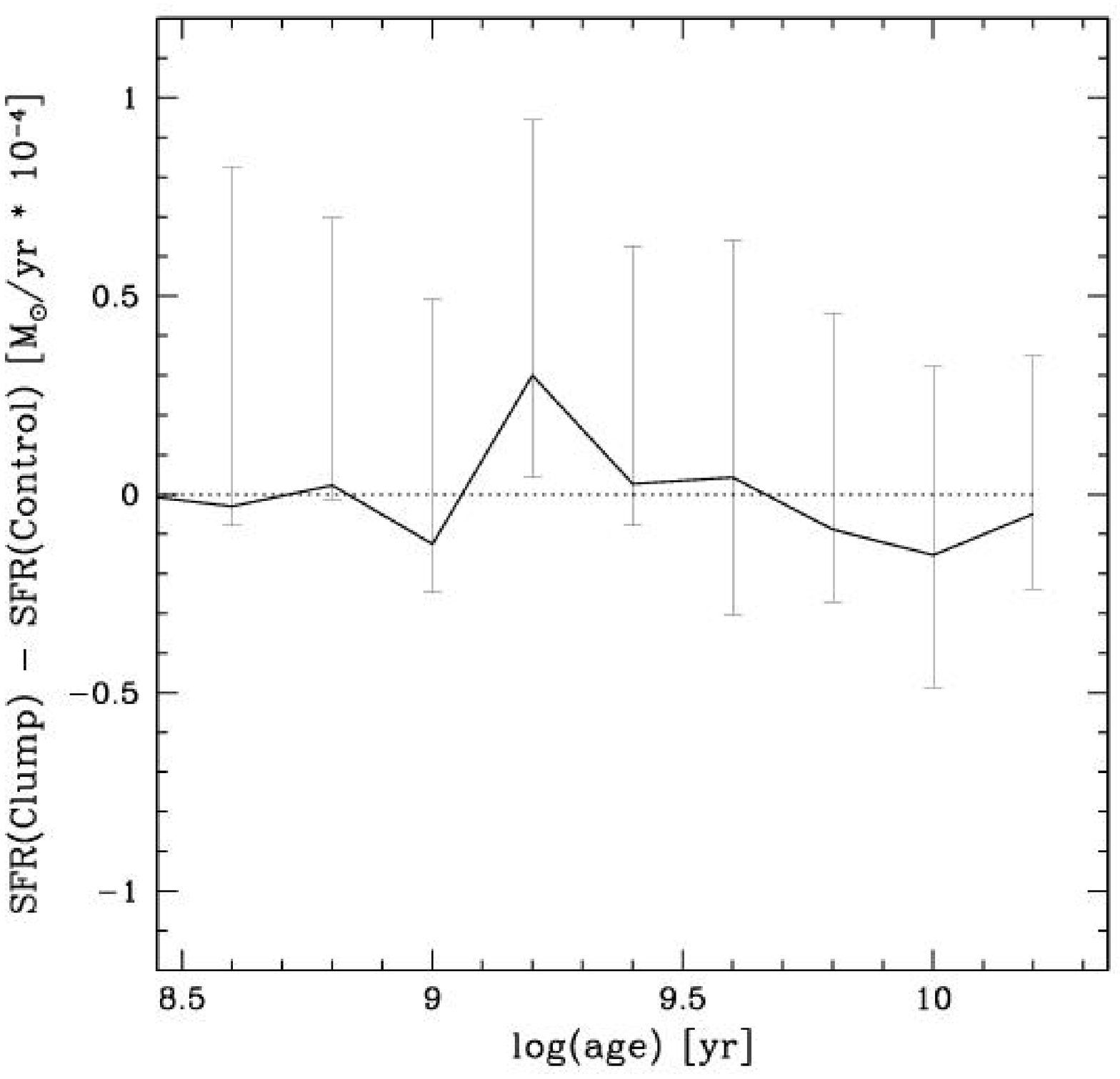}
   \caption{\label{fig:sfh_comp} \scriptsize 
The two star-formation histories in Figure 4 are now subtracted
after scaling the control field.
We show the clump minus the control star formation histories
with composite error bars. The most significant difference
between the two histories is at log(age)$=$9.2. Note that,
as described in the text, the CMD is binned coarsely 
to reduce shot noise in the analysis. Short burst of star
formation are thus reduced in emphasis by StarFISH, and
are best seen in the actual CMD.}
 \end{center}
\end{figure}

Figure 6 overlays four isochrones on the CMD of the clump field: three
with z=0.004 and ages of log(age)=9.1, 9.2. 9.3, and one with
log(age)$=$10.0 and z=0.001.  StarFISH estimates the star formation in
bins of 0.2 in log(t), so the 9.2 bin in StarFISH encompasses the 9.1,
9.2, and 9.3 isochrones.  Interpolating by eye, a good fit to the
younger main sequence would be log(age)=9.15. Since this one isochrone
effectively fits the younger main sequence, we can say with confidence
that the burst was short, perhaps only 300 Myr in duration.

\begin{figure}
 \begin{center}
  \plotone{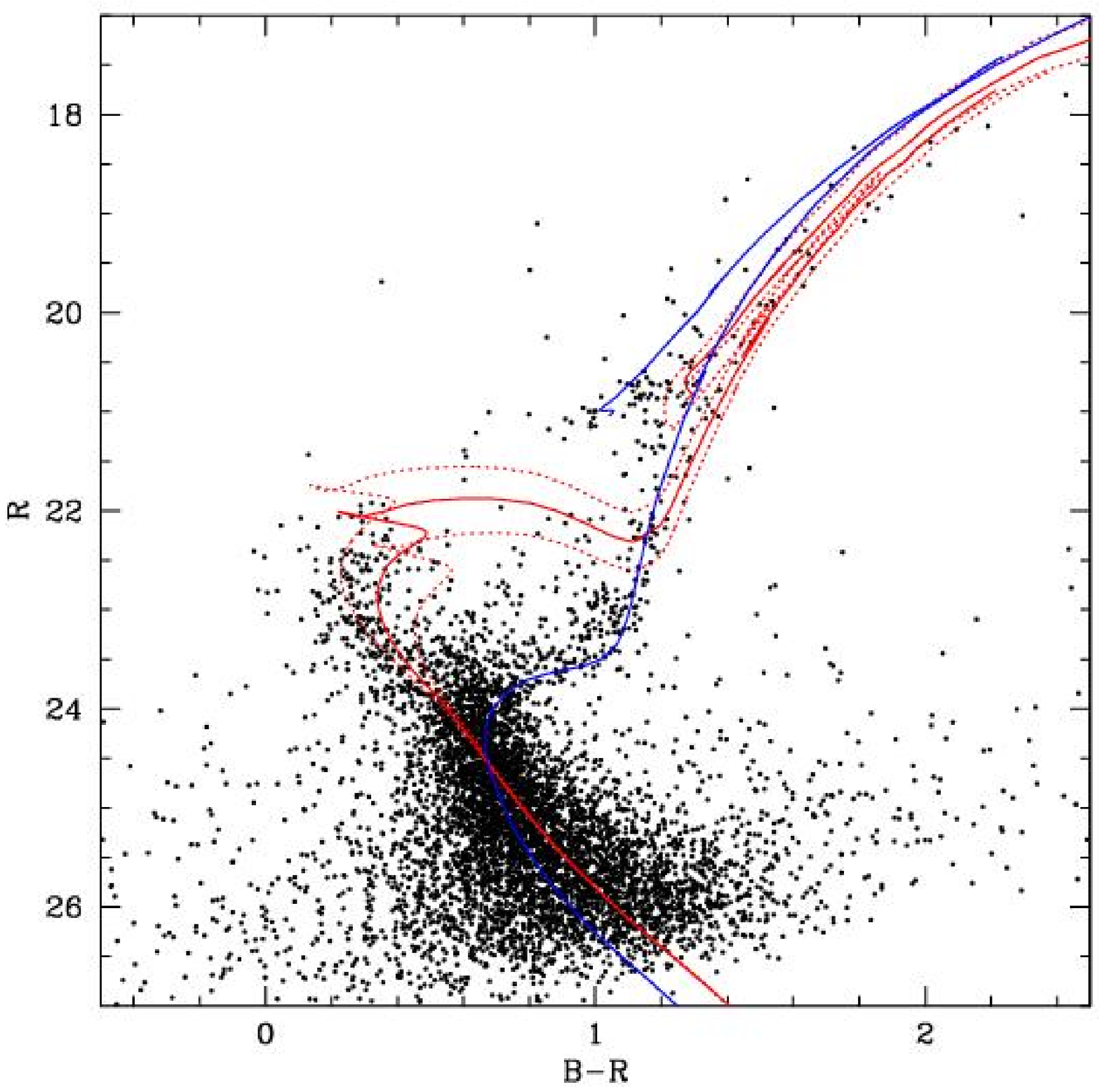}
   \caption{\label{fig:sfh_iso} \scriptsize Fornax clump field
with overlaid Girardi et al.~2002 isochones. Isochrones with
abundance z$=$0.004 and log(age)=9.1, 9.2 (solid line), 9.3, 
and z$=$0.001 and log(age)$=$10.0, are plotted.
The effects of a reddening of E(B-V)$=$0.02 have been applied to the isochrones.}
 \end{center}
\end{figure}

The nature of the additional population in the clump can also be
seen by statistically subtracting the scaled control CMD from
that of the clump.   To do this,  we first defined a grid of cells in the
CMD in regular intervals of 0.15 and 0.5 mag in $B-R$ color and $R$
magnitude, respectively.  Within each bin, completeness corrections
from the false-star measurements were applied.   As noted, the control
field has more
stars than the clump field.  
A scaling factor of 1.4, slightly different from the 1.5 derived from simple star
counts above, gave the best results.
With this scaling, we then statistically removed
stars in the clump field located within a small distance (similar
to the bin sizes) of stars selected at random from the control CMD.
The method works reasonably well near the main sequence, but, due to the
small numbers of stars, less effectively on the SGB and RGB.  Note too
that the method is quite sensitive to errors in the completeness
corrections; for this reason we only trust the subtracted CMD as a
useful diagnostic of the clump population for $R \leq 24.5$.

\begin{figure}
 \begin{center}
  \includegraphics[scale=0.7, angle=270]{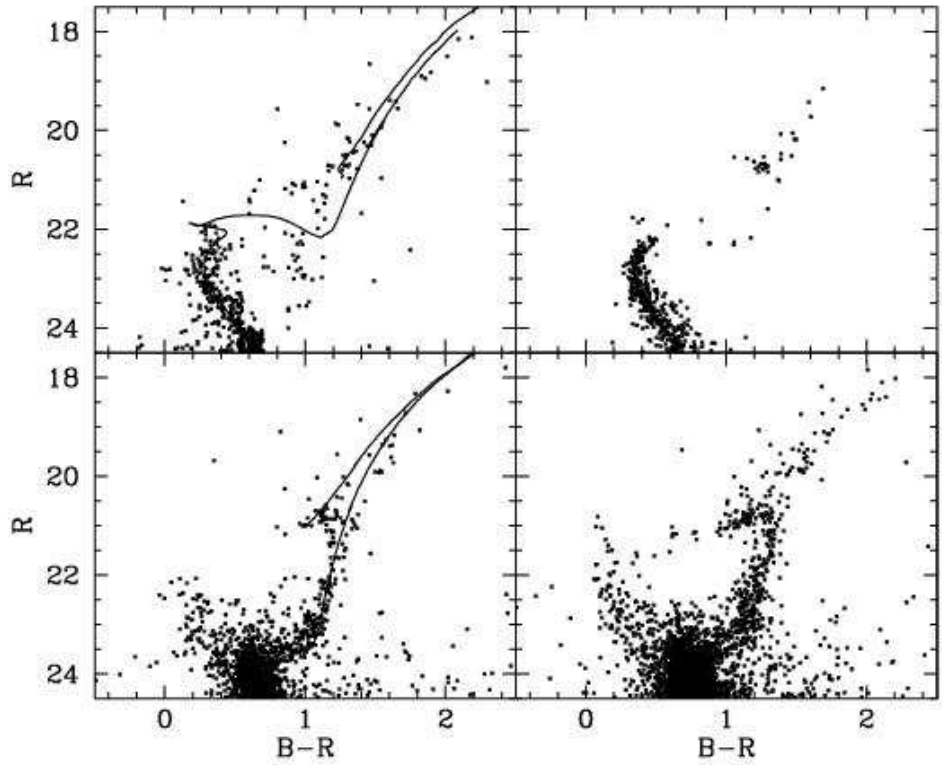}
   \caption{\label{fig:sfh_diff} \scriptsize The difference between the clump
field and the control field is shown here after statistically subtracting
the scaled control field from the clump field. This statistical subtraction
was only done to R=24.5 mag. Upper left: the residual
clump field CMD after subtracting the scaled control CMD as
described in the text. This is the 1.4 Gyr population, or ``the burst of
star formation in this clump field.''
The 1.4 Gyr isochrone with z$=$0.004 is also overplotted.
Upper right: a synthetic 1.4 Gyr population CMD with the same number of
main-sequence stars as in the upper left diagram. Lower left: the CMD of the clump population
that was actually removed from the raw clump CMD to make
the upper left diagram. This is therefore
the ``typical part of Fornax'' in the clump field. The 10 Gyr isochrone from the previous
figure is overlaid. Lower right: the
original control field CMD.}
 \end{center}
\end{figure}

Figure 7 shows the results of this subtraction. The upper-left panel is the CMD of the clump
that is different from the scaled control field. The lower-left CMD is what was actually
subtracted from the left side of Figure 3, in other words the CMD of the clump
that is statistically identical to the scaled and binned control CMD.
The 1.4 Gyr Padua isochrone for z=0.004 is overplotted at upper left, as is the 10 Gyr
isochrone for z=0.001 at lower left. At the upper right we show a synthetic CMD
from the StarFISH solution at 9.1$<$log(age)$<$9.3. Finally, the control CMD is shown in the lower
right.

The good fit of the 1.4 Gyr isochrone to the subtracted CMD is our final way of
arguing that this extra population can be called a burst of star formation. 
While indeed there are many places in Fornax with some young stars, this
density clump is different from the control field in this one major way,
the addition of a substantial number of younger stars.
The characteristics of this
extra population, with $z=0.004$, imply that the gas
making these stars was significantly pre-enriched. A small
galaxy such as Ursa Minor would be expected to have a metallicity
of roughly $z=0.0002$, [Fe/H] roughly $-2.0$, which is lower
than the metallicity of the oldest substantial population
in Fornax. Could this gas have been ``Fornax gas''?
Figure 8 shows (left panel) Padua isochrones for an age of 1.4 Gyr
and metallicities of z$=$ 0.002, 0.004, and 0.008 ([Fe/H]$=$$-1.0$, $-0.7$, and $-0.4$). The more
metal-poor isochrone is not a good fit at this age. If we adopt [Fe/H]$=$$-1.0$
(Tolstoy et al.~(2001)
and allow the derived age to change to 1.8 Gyr, we get the right panel in
Figure 8. This isochrone does not fit the main sequence very well,
and fits the giants quite badly. For these reasons, StarFISH
found only a very small metal poor population at this age, which
is now confirmed by the isochrone fits to the subtracted CMD.
To the extent that the Padua isochrones are on the same metallicity scale
as the spectroscopic measurements, we conclude that the typical
red giant in Fornax is more metal poor than the stars in the 1.4 Gyr burst. 
The most logical place to make this enriched gas is in Fornax itself.

\begin{figure}
 \begin{center}
 \plotone{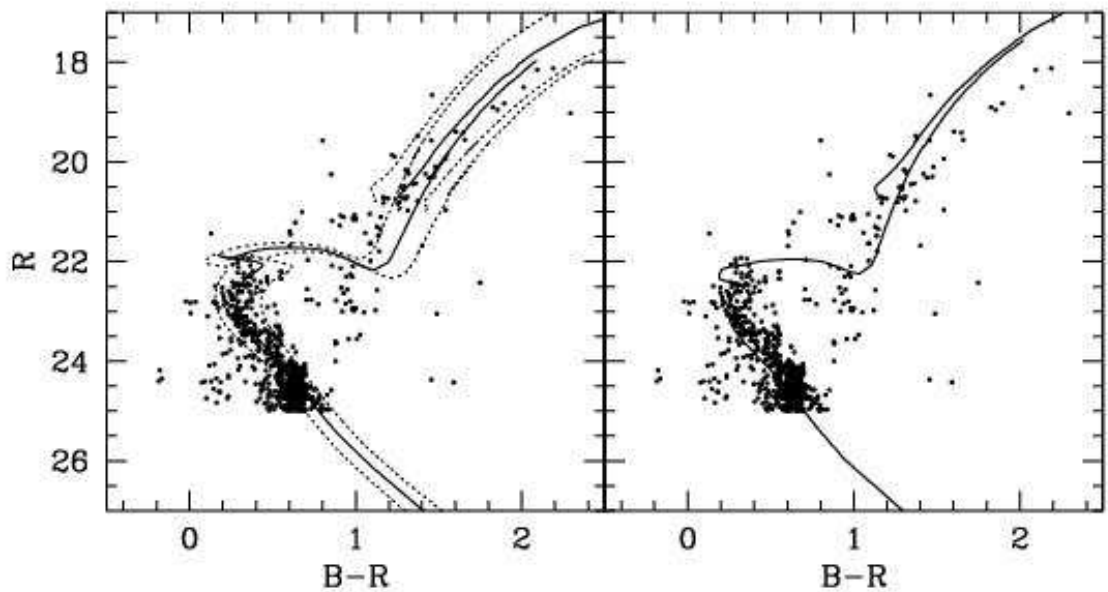}
   \caption{\label{fig:2pane} \scriptsize Fits of different metallicity isochrones
to the young population. Left Panel: Padua isochrones at an age of 1.4 Gyr with metallicities
of z$=$ 0.002, 0.004, and 0.008 ([Fe/H]=-1.0, -0.7, and -0.4. z$=$0.004 is the best fit.
The solid curve is the isochrone for z$=$0.004.
Right Panel: Padua isochrone with age of 1.8 Gyr at z$=$0.002. This isochone, with
the average spectroscopic metallicity of the Fornax red giants, is not a good fit to this
younger population. }
 \end{center}
\end{figure}

The next important observation
of this field would be the kinematics of stars and detailed chemical
compositions in this small
area of Fornax, to see if star formation history is the only major
difference in this field.
Of course, the questions of how Fornax retained
gas until 1.4 Gyr ago, or gained gas, and why Fornax did not undergo a global burst
of this intensity, and why the clump has not dynamically
mixed, remain important unsolved questions.

We thank the staff at Las Campanas and Magellan for making
observing at LCO a productive pleasure. 
EO is partially supported by NSF grants AST-0098518, 0205790, and 0507511.
MM and MW are partially supported by NSF grants AST-0098661 and AST-0206081
and AST-05xxxxxx.
JH is supported by NASA through Hubble Fellowship grant HF-01160.01-A
awarded by the Space Telescope Science Institute, which is operated
by the Association of Universities for Research in Astronomy, Inc.,
under NASA contract NAS 5-26555.
GDC and MC are supported in part by
funding from the Australian Research Council through Discovery
Projects Grants DP0343156 and DP0343508.

\end{document}